\documentclass[conference]{IEEEtran}
\IEEEoverridecommandlockouts
\usepackage{amsmath,amsfonts,amssymb}
\usepackage{algorithm}
\usepackage{algorithmic}
\usepackage{array}
\usepackage{hyperref}
\usepackage{textcomp}
\usepackage{stfloats}
\usepackage{url}
\usepackage{verbatim}
\usepackage{graphicx}
\usepackage{subfigure}
\usepackage{cite}
\usepackage{xurl}
\usepackage{multirow}
\usepackage{cleveref}

\crefname{equation}{}{}
\usepackage{tikz}  
\newcommand*\circled[1]{\tikz[baseline=(char.base)]{
            \node[shape=circle,draw,inner sep=1pt] (char) {#1};}}

\begin{document}

\title{Safe Decentralized Operation of EV Virtual Power Plant with Limited Network Visibility via Multi-Agent Reinforcement Learning}

\author{\IEEEauthorblockN{Chenghao Huang\textsuperscript{1}, Jiarong Fan\textsuperscript{1}, Weiqing Wang\textsuperscript{1}, Hao Wang\textsuperscript{1,2*}}
\IEEEauthorblockA{\textsuperscript{1}Department of Data Science and AI, Faculty of IT, Monash University, Australia \\
\textsuperscript{2}Monash Energy Institute, Monash University, Australia\\
}
\thanks{*Corresponding author: Hao Wang (hao.wang2@monash.edu).}
\thanks{This work was supported in part by the Australian Research Council (ARC) Discovery Early Career Researcher Award (DECRA) under Grant DE230100046.}
}

\maketitle

\begin{abstract}
As power systems advance toward net-zero targets, behind-the-meter renewables are driving rapid growth in distributed energy resources (DERs). Virtual power plants (VPPs) increasingly coordinate these resources to support power distribution network (PDN) operation, with EV charging stations (EVCSs) emerging as a key asset due to their strong impact on local voltages.
However, in practice, VPPs must make operational decisions with only partial visibility of PDN states, relying on limited, aggregated information shared by the distribution system operator.
This work proposes a safety-enhanced VPP framework for coordinating multiple EVCSs under such realistic information constraints to ensure voltage security while maintaining economic operation. We develop Transformer-assisted Lagrangian Multi-Agent Proximal Policy Optimization (TL-MAPPO), in which EVCS agents learn decentralized charging policies via centralized training with Lagrangian regularization to enforce voltage and demand-satisfaction constraints.
A transformer-based embedding layer deployed on each EVCS agent captures temporal correlations among prices, loads, and charging demand to improve decision quality.
Experiments on a realistic 33-bus PDN show that the proposed framework reduces voltage violations by approximately 45\% and operational costs by approximately 10\% compared to representative multi-agent DRL baselines, highlighting its potential for practical VPP deployment.

\end{abstract}


\section{Introduction}

Decarbonization commitments and the global push toward net-zero energy systems are accelerating the integration of renewable resources into distribution networks \cite{IEAnetzero}. As behind-the-meter PV adoption increases, distributed energy resources (DERs), such as rooftop PV, residential batteries, and electric vehicles (EVs) \cite{IEA2024}, are increasingly coordinated through virtual power plants (VPPs) to provide flexibility services and support grid stability \cite{pudjianto2007virtual}.

However, the rapid growth of DERs introduces significant operational challenges for power distribution networks (PDNs). High penetrations of distributed PV can cause reverse power flows and voltage rise, while uncoordinated EV charging can create peak load surges and voltage violations \cite{9536577}. Among various DERs, EV charging stations (EVCSs) stand out as a priority for VPP coordination because their high-throughput, long-dwell, and highly flexible yet spatially concentrated charging demand can induce substantial voltage deviations in PDNs.

Recent research on EV charging coordination has ranged from centralized rule-based control and optimization \cite{10142097,9591485} to multi-agent reinforcement learning (MARL) approaches that improve scalability and adaptability under uncertainty \cite{10312315,10557451}. Ensuring grid safety, however, remains challenging. Existing efforts include embedding operational limits in the action space \cite{10143999}, reward shaping to penalize constraint violations \cite{10265006,10960355}, and Lagrangian regularization for principled constraint handling \cite{stooke2020responsive,zhang2023safe}. Despite the progress, safe mechanisms within MARL remain relatively underexplored in power system applications.

Furthermore, VPPs seldom have full access to PDN states or topology due to privacy, regulatory, and cybersecurity constraints. This limited visibility makes it difficult to anticipate how local charging actions propagate through electrical coupling to affect voltages and network constraints elsewhere, leading to potentially unsafe or inefficient decisions.
Therefore, two key challenges remain insufficiently addressed in existing MARL-based EVCS coordination: limited safety guarantees during learning and deployment, which hinder principled voltage-limit enforcement, and the common assumption of full grid-state visibility, whereas real VPP-DSO interactions offer only partial, privacy-preserving indicators. These limitations restrict the practical deployment of MARL in real distribution networks.

In this work, we study the problem of coordinating multiple EVCSs under the realistic constraint that a VPP only receives partial and privacy-preserving grid indicators from the DSO. We integrate transformer-based temporal capturing with multi-agent reinforcement learning and principled safety regularization, forming a framework tailored for EVCS coordination under partial PDN visibility.
The main contributions of this work are summarized as follows:
\begin{itemize}
    \item We formalize a realistic VPP-DSO coordination setting where multiple EVCSs operate under partial PDN visibility, capturing core challenges in achieving economic charging and voltage safety amid uncertainties in EV demand, PV output, and electricity prices.
    \item We propose TL-MAPPO, a safety-enhanced multi-agent RL framework that learns decentralized EVCS policies under centralized training. It integrates Lagrangian regularization for enforcing safety constraints and a transformer-based embedding layer for improved temporal context under limited grid visibility.
    \item Experiments on the IEEE 33-bus system show that TL-MAPPO reduces voltage violations by about 45\% and operational costs by about 10\% compared with representative state-of-the-art MARL baselines, demonstrating its potential for safe and practical EVCS coordination.
\end{itemize}

\section{System Model}\label{Sec:SM}

In a PDN operated by a DSO, a VPP coordinates multiple EVCSs deployed at different electric buses.
The VPP manages EV charging activities and rooftop photovoltaic (PV) usage to minimize operational costs while maintaining voltage safety and fulfilling user demand requirements.
An overview of the system configuration is shown in Fig. \ref{fig:system model}.
Detailed PDN information (e.g., voltage and power at nodes) is generally available to the VPP, based on pilot projects \cite{ARENA}, and we \textbf{assume} that the DSO grants each VPP-managed EVCS access to partial PDN information corresponding to its local neighborhood, including the voltage magnitudes and aggregated loads of its hosting bus and adjacent buses within one hop.
This limited visibility reflects realistic communication and privacy constraints between the DSO and the VPP. 

\subsection{PDN}
The PDN consists of $N$ buses connected through distribution lines and operates over a discrete time horizon $\mathcal{T}={1,2,\dots,T}$.
At each bus $i\in\mathcal{N}$ and time step $t\in\mathcal{T}$, the active power injection is denoted by $p_{i,t}$, and the voltage magnitude by $v_{i,t}$.
Give the allowable voltage magnitude range, the voltage of node $i$ is constrained as
\begin{equation}
V^{\text{min}} \le v_{i,t} \le V^{\text{max}}, \quad \forall i \in \mathcal{N}, \, t \in \mathcal{T}.
\end{equation}

\subsection{EVCS and Neighborhood Nodes}
The VPP manages a set of EVCSs $\mathcal{K}={1,\dots,K}$, each equipped with rooftop PV generation and a set of chargers $\mathcal{C}_k={1,\dots,C_k}$.
Each EVCS $k$ is physically connected to one bus of the PDN, represented by a binary deployment matrix $\mathbf{K}\in\{0,1\}^{N\times K}$, where $K_{ik}=1$ if EVCS $k$ is located at bus $i$, and $K_{ik}=0$ otherwise,
\begin{align}
\sum_{k\in\mathcal{K}} K_{ik} \le 1, \quad \forall i\in\mathcal{N}.
\end{align}

For each bus $i \in \mathcal{N}$, let $\mathcal{N}_i^{(1)}$ denote its 1-hop neighbor set, which contains all buses directly connected to bus $i$ through distribution lines:
\begin{align}
\mathcal{N}_i^{(1)} = \{ j \in \mathcal{N} \setminus \{i\} \mid (i,j) \in \mathcal{L} \},
\end{align}
where $\mathcal{L}$ is the set of distribution lines connecting pairs of buses.

Given the EVCS deployment matrix $\mathbf{K}\in\{0,1\}^{N\times K}$,
the 1-hop neighbor bus set associated with EVCS $k$ (located at bus $i$ where $K_{ik}=1$) is defined as
\begin{align}
\mathcal{N}_k^{(1)} = \mathcal{N}_i^{(1)} = \{ j \in \mathcal{N} \setminus \{i\} \mid (i,j) \in \mathcal{L}, K_{ik}=1 \}.
\end{align}
This set represents all the buses physically adjacent to the bus hosting EVCS $k$, whose voltage and injection information can be partially accessed under the DSO-VPP collaboration assumption.

On the EV side, each charger must also respect the physical charging requirements imposed by individual EVs, characterized by their arrival-departure windows and target SoC trajectories.
Specifically, $\text{SoC}^{\text{target}}_{c_k}(t)$ is the expected SoC trajectory between the arrival and departure times of each EV:
\begin{align}
\text{SoC}^{\text{target}}_{c_k}(t)=
\text{SoC}^{\text{arr}}_{c_k}+
\frac{t-T^{\text{arr}}_{c_k}}{T^{\text{dep}}_{c_k}-T^{\text{arr}}_{c_k}}
\big(\text{SoC}^{\text{dep}}_{c_k}-\text{SoC}^{\text{arr}}_{c_k}\big).
\end{align}

Each charger satisfies the operational constraints below:
\begin{subequations}\label{eq:charger limit}
\begin{align}
&0\le p_{c_k}^{\text{ch}}(t)\le P_{c_k}^{\text{ch,max}},\quad
0\le p_{c_k}^{\text{dis}}(t)\le P_{c_k}^{\text{dis,max}},\\
&p_{c_k}^{\text{ch}}(t)p_{c_k}^{\text{dis}}(t)=0,\\
&\text{SoC}_{c_k}^{\text{min}}\le\text{SoC}_{c_k}(t)\le\text{SoC}_{c_k}^{\text{max}},\\
&\Delta\text{SoC}_{c_k}(t)=
\eta^{\text{ch}}p_{c_k}^{\text{ch}}(t)-\tfrac{1}{\eta^{\text{dis}}}p_{c_k}^{\text{dis}}(t).
\end{align}
\end{subequations}

\begin{figure}[!t]
\centering
\includegraphics[width=0.4\textwidth]{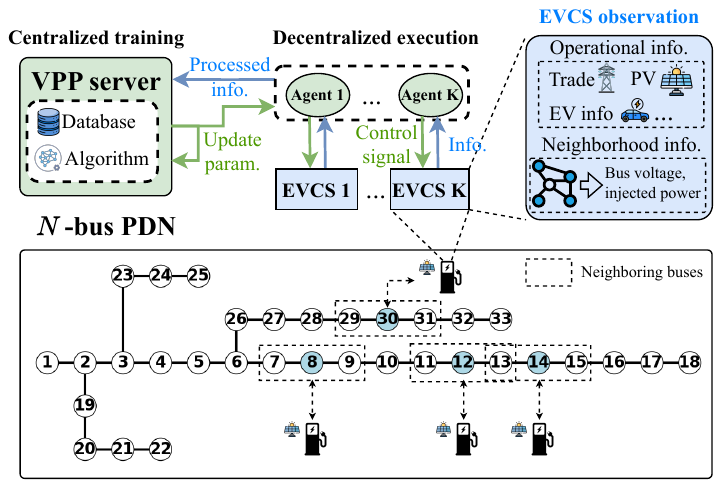}
\caption{System model of decentralized EVCS coordination under a VPP's centralized training in a PDN.} 
\vspace{-10pt}
\label{fig:system model}
\end{figure}

\subsection{Objective}
The objective of the VPP is to coordinate EVCS operations to minimize overall costs while satisfying physical and operational constraints.  
At each time $t$, the $k$-th EVCS controls the charging and discharging power of its chargers, $\{p_{c_k}^{\text{ch}}(t), p_{c_k}^{\text{dis}}(t)\}_{c_k=1}^{C_k}$, subject to charger and battery limits.  
The optimization objective is formulated as:
\begingroup
\footnotesize
\begin{equation}
\min_{\mathbf{p}^{\text{ch}},\mathbf{p}^{\text{dis}}}
\sum_{t\in\mathcal{T}}\Bigg[\sum_{k\in\mathcal{K}}\big[\beta_1 f_k^{\text{TR}}(t)+\beta_2 f_k^{\text{DG}}(t) + \beta_3 f_k^{\text{DS}}(t) \big] + \beta_4 f^{\text{VT}}(t)\Bigg],
\label{eq:objective}
\end{equation}
\endgroup
where $f_k^{\text{TD}}(t)$ and $f_k^{\text{DG}}(t)$ denote the energy trading and battery degradation costs (also called cycling overhead in Section \ref{Sec:E}), $f^{\text{VT}}(t)$ quantifies total voltage violations, $f_k^{\text{DS}}(t)$ represents the charging dissatisfaction penalty of EVCS $k$, and $\beta_1 - \beta_4$ are the coefficients of each term.
Specifically, the cost components are defined as
\begin{align}
&f_k^{\text{TD}}(t)=
\begin{cases}
\lambda^{\text{buy}}_t p_k^{\text{TD}}(t), & p_k^{\text{TD}}(t)>0,\\
\lambda^{\text{sell}}_t p_k^{\text{TD}}(t), & \text{otherwise},
\end{cases} \\
&f_k^{\text{DG}}(t)=
\alpha_e\sum_{c_k\in\mathcal{C}_k}\big([p_{c_k}^{\text{ch}}(t)]^2+[p_{c_k}^{\text{dis}}(t)]^2\big),\\
&f^{\text{VT}}(t)=
\sum_{i\in\mathcal{N}}\big([v_{i,t}-V^{\text{max}}]^+ + [V^{\text{min}}-v_{i,t}]^+\big),\\
&f_k^{\text{DS}}(t)=
\sum_{c_k\in\mathcal{C}_k}
\big[\text{SoC}^{\text{target}}_{c_k}(t)-\text{SoC}_{c_k}(t)\big]^+,
\end{align}
where $[x]^+=\max(0,x)$, and $\text{SoC}^{\text{target}}_{c_k}(t)$ is the EV $c_k$'s target SoC at time $t$, linearly interpolated from arrival to departure.
The power traded between the EVCS $k$ and the grid is
\begin{equation}
p_k^{\text{TD}}(t)=
\sum_{c_k\in\mathcal{C}_k}\big[p_{c_k}^{\text{ch}}(t)-p_{c_k}^{\text{dis}}(t)\big]-p_{k,t}^{\text{PV}}.
\end{equation}

\begin{figure*}[!t]
\centering
\includegraphics[width=0.85\textwidth]{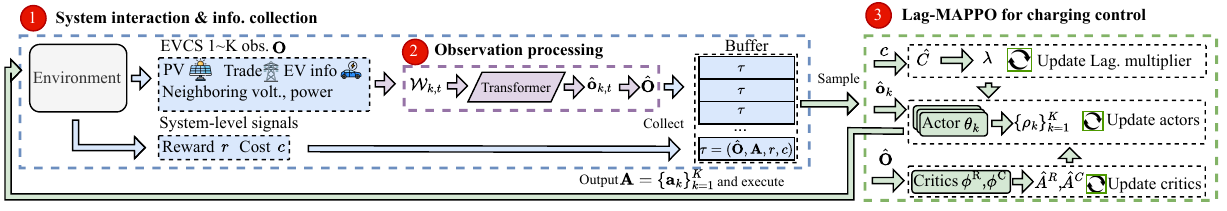}
\caption{The developed TL-MAPPO for safe EVCS coordination under partial PDN visibility, primarily built on MARL and Transformer.} 
\vspace{-15pt}
\label{fig:methodology}
\end{figure*}

\section{Methodology}\label{Sec:M}
This section presents the overall framework of the proposed TL-MAPPO method, illustrated in Fig. \ref{fig:methodology}, which integrates three tightly coupled methodological components for safe EVCS coordination under partial network visibility. First, the EVCS coordination problem is formulated as a partially observable constrained Markov decision process (PO-CMDP), which captures long-horizon decision making with limited local observations and safety constraints. Based on this formulation, each EVCS agent collects local measurements and operational information to construct a temporal observation sequence. Second, a Transformer-based observation encoder processes this sequence to extract compact temporal representations, which serve as the input to decentralized control policies. Third, these encoded observations are used by a Lagrangian MAPPO algorithm, where decentralized actors are trained with centralized critics to jointly optimize economic performance and enforce voltage and demand-satisfaction constraints. During execution, each EVCS operates independently using its local policy, while safety is ensured through the learned Lagrangian regularization.

\subsection{PO-CMDP Formulation}
The coordination of EVCSs under partial visibility and safety constraints is modeled as a PO-CMDP
$\mathcal{M}=\langle\mathcal{K},\mathcal{S},\mathcal{O},\mathcal{A},\mathbb{P},\mathbb{R},\mathbb{C},\gamma\rangle$,
where $\mathcal{K}$ is the set of EVCS agents, $\mathcal{S}$ the global state, $\mathcal{O}=\prod_k\mathcal{O}_k$ the joint observations, $\mathcal{A}=\prod_k\mathcal{A}_k$ the joint actions, $\mathbb{P}$ the transition, $\mathbb{R}$ the reward, $\mathbb{C}$ the safety cost, and $\gamma$ the discount factor. At time $t$, EVCS $k$ observes
\begin{align}
\mathbf{o}_k(t)=\Big[ & \underbrace{[p_{i,t}, v_{i,t}]_{i\in\mathcal{N}_k^{(1)}}}_{\text{Neighborhood}}, \underbrace{p_{k,t}^{\text{PV}},\lambda^{\text{buy}}_t,\lambda^{\text{sell}}_t}_{\text{PV and prices}},\notag\\
&\underbrace{[T^{\text{dep}}_{c_k}-t,\text{SoC}^{\text{dep}}_{c_k}-\text{SoC}_{c_k}]_{c_k\in\mathcal{C}_k}}_{\text{EV info}}\Big].
\end{align}
The action vector $\mathbf{a}_k(t)=[a_{c_k}(t)]_{c_k=1}^{C_k}$ denotes charging/discharging power, bounded by $-P^{\text{dis,max}}_{c_k}\le a_{c_k}(t)\le P^{\text{ch,max}}_{c_k}$ and subject to SoC/exclusivity constraints. The reward follows the economic objective:
\begin{equation}
r(t)=\sum_{k\in\mathcal{K}}\big[\beta_1 f_k^{\text{TD}}(t)+\beta_2 f_k^{\text{DG}}(t)\big].
\end{equation}
The safety cost aggregates voltage violations and demand dissatisfaction:
\begin{align}
c(t)=\beta_3 \sum_{k\in\mathcal{K}}f_k^{\text{DS}}(t) + \beta_4 f^{\text{VT}}(t).
\end{align}

\subsection{Transformer-based Observation Processing}\label{sec:transformer}
We adopt a Transformer \cite{vaswani2017attention} to capture temporal correlations among the input features. For each EVCS $k$, an observation window $\mathcal{W}_{k}(t)=[\mathbf{o}_{k}(t-t'+1);\ldots;\mathbf{o}_{k}(t)]$ is encoded by a multi-head self-attention encoder:
\begin{align}
\mathbf{h}_k^{\text{Trans}}(t)=g^{\text{Trans}}_k\big(\mathcal{W}_{k,t}\big),
\end{align}
where $g_k^{\text{Trans}}(\cdot)$ denotes the Transformer encoder that extracts temporal representations from the observation window $\mathcal{W}_{k,t}$.
Next, the processed observation fed to MARL is $\hat{\mathbf{o}}_k(t)=[\mathbf{h}_k^{\text{Trans}}(t)]$, with $\hat{\mathbf{O}}(t)=[\hat{\mathbf{o}}_1(t),...,\hat{\mathbf{o}}_K(t)]$.

\subsection{Lag-MAPPO for EVCS Coordination}\label{sec:lagmppo}
We adopt a Lag-MAPPO framework, where decentralized actors $\{\pi_{\theta_k}\}$ are trained with two centralized critics $V_{\phi^R}$ and $V_{\phi^C}$. Each EVCS agent samples its control action from
\begin{equation}
\mathbf{a}_k(t)\sim\pi_{\theta_k}(\mathbf{a}_k(t)|\hat{\mathbf{o}}_k(t)).
\end{equation}
Given trajectories $\tau(t)=[\hat{\mathbf{O}}(t),\mathbf{A}(t),r(t),c(t)]$, the normalized advantages for reward and cost are
\begin{subequations}
\begin{align}
    &\hat{A}^R(t)=\sum_{l=0}^{T-t}\gamma^{l}r(t+l)-V_{\phi^R}\big(\hat{\mathbf{O}}(t)\big),\\
    &\hat{A}^C(t)=\sum_{l=0}^{T-t}\gamma^{l}c(t+l)-V_{\phi^C}\big(\hat{\mathbf{O}}(t)\big),
\end{align}
\end{subequations}
where $V_{\phi^R}$ and $V_{\phi^C}$ estimate the expected reward and cost under current policy, respectively.

The actor optimizes the clipped PPO objective with Lagrangian shaping:
\begingroup
\footnotesize
\begin{subequations}
\begin{align}
&\hat{A}^{\text{Lag}}(t)=\hat{A}^R(t)-\lambda\,\hat{A}^C(t),\\
&L^{\text{Act}}_k=\mathbb{E}\Big[-\min\{\rho_k(t),\sigma[\rho_k(t),1-\epsilon,1+\epsilon]\}\cdot\hat{A}^{\text{Lag}}(t)\Big],\\
&\rho_k(t)=\frac{\pi_{\theta_k}(\mathbf{a}_k(t)|\hat{\mathbf{o}}_k(t))}{\pi_{\theta'_k}(\mathbf{a}_k(t)|\hat{\mathbf{o}}_k(t))},
\end{align}
\end{subequations}
\endgroup
where $\rho_k(t)$ is the importance ratio between current and old policies, $\sigma$ is clipping operation, $\epsilon$ is the PPO clipping threshold, and $\lambda$ is the Lagrangian multiplier.

The critic minimizes the squared error of the Lagrangian-shaped return:
\begin{equation}
L^{\text{Critic}}=\mathbb{E}\Big[\Big(V_{\phi^C}\big(\hat{\mathbf{O}}(t)\big)-\big[r(t)-\lambda\,c(t)\big]\Big)^2\Big].
\end{equation}

The multiplier $\lambda$ is updated by projected dual ascent:
\begin{equation}
\lambda\leftarrow\big[\lambda+\eta^{\text{Lag}}(\hat{C}(t)-\bar{C})\big]^+,
\end{equation}
where $\eta^{\text{Lag}}$ is the dual learning rate, $\hat{C}(t)$ is the observed cost, $\bar{C}$ is the desired constraint threshold, and $[\cdot]^+$ ensures non-negativity.  
Finally, parameters $\theta_k$, $\phi^R$, and $\phi^C$ are updated by gradient descent on $L^{\text{Act}}_k$ and $L^{\text{Critic}}$, respectively.  
This yields stable policy improvement while prioritizing voltage and demand safety under partial visibility.

\section{Simulations}\label{Sec:E}

\subsection{Experimental Setup}

Four EVCSs, each equipped with ten chargers, are deployed at buses 8, 12, 14, and 30, as shown in Fig. \ref{fig:system model}.
The simulation adopts a 33-bus PDN operated over a one-day horizon (288 steps of 5 minutes each), with EV SoC constrained within [0,80] kWh, charging/discharging rates within [-22,22] kW, and efficiency of 0.95.
The network voltage is maintained within the safety range of [0.95,1.05] p.u.

For the numerical simulation, we use daily EV data collected by Caltech \cite{lee2019acn} and rooftop PV generation data of Solar Home Electricity Dataset \cite{ratnam2017residential} provided by Ausgrid's electricity network from July 1, 2010, to June 30, 2013.
The wholesale price of AEMO is used for energy transaction \cite{aemo_dashboard}, and the TOU price is set by the DSO, taking Melbourne as an example. 
% In Fig. \ref{fig:ev_stat}, 
In terms of the EV data, the features of over 900,000 EVs with charging duration shorter than one day (1440 minutes) are included, where most of the EV demands are less than 20 kWh in their charging durations primarily ranged from 10 to 800 minutes.
All EVs follow a First-In-First-Out principle.

In terms of the comparison baselines, we evaluate our developed TL-MAPPO against three DRL baselines: \begin{footnotesize}\circled{1}\end{footnotesize} MAPPO - A centralized-training, decentralized-execution extension of PPO, with good stability and strong performance.
\begin{footnotesize}\circled{2}\end{footnotesize} MATD3 - A multi-agent version of TD3 that mitigates overestimation bias and is effective for continuous control under deterministic policies.
\begin{footnotesize}\circled{3}\end{footnotesize} MASAC - A multi-agent extension of SAC that leverages entropy regularization to enhance exploration and improve learning robustness.

To comprehensively evaluate the performance of the experimented method, we report four key metrics that reflect economic operation, constraint satisfaction, and service quality:
\begin{footnotesize}\circled{1}\end{footnotesize} \textbf{Energy cost} (\$): Total operational cost, equal to the cumulative reward over one day.
\begin{footnotesize}\circled{2}\end{footnotesize} \textbf{Cycling overhead} (unitless): Ratio of total energy throughput (charge + discharge) to charging demand.
\begin{footnotesize}\circled{3}\end{footnotesize} \textbf{Average voltage violation} (p.u./5 min): Mean per-step voltage violation over all buses.
\begin{footnotesize}\circled{4}\end{footnotesize} \textbf{Average demand dissatisfaction} (kWh/EV): Mean unmet charging demand per EV over one episode.

\subsection{Result Analysis}

\begin{table}[t]
\centering
\scriptsize
\caption{Test performance over 100 independent episodes.}
\vskip -6pt
\renewcommand\arraystretch{1.2}
\setlength{\tabcolsep}{0.8mm}{
\begin{tabular}{lp{1.33cm}p{1.5cm}p{1.8cm}p{1.85cm}}
\hline\hline
\textbf{Method} & \textbf{Energy cost (\$)} & \textbf{Cycling overhead} & \textbf{Avg. volt. vio. (p.u./5 min)} & \textbf{Avg. demand dissat. (kWh/EV)} \\ \hline
MATD3 & 144.2 $\pm$ 3.8 & 118.7 $\pm$ 2.5\% & 7.8 $\pm$ 0.6$\times10^{-3}$ & 0.88 $\pm$ 0.07 \\
MAPPO & 140.6 $\pm$ 4.1 & 122.4 $\pm$ 3.2\% & 6.1 $\pm$ 0.5$\times10^{-3}$ & 0.76 $\pm$ 0.06 \\
MASAC & 148.9 $\pm$ 3.9 & 126.5 $\pm$ 2.9\% & 9.3 $\pm$ 0.8$\times10^{-3}$ & 1.12 $\pm$ 0.09 \\
\textbf{TL-MAPPO} & \textbf{133.5 $\pm$ 3.4} & \textbf{110.2 $\pm$ 2.1\%} & \textbf{4.2 $\pm$ 0.4}$\times10^{-3}$ & \textbf{0.58 $\pm$ 0.05} \\
\hline\hline
\end{tabular}
\vspace{-10pt}
\label{Tab:test_results}}
\end{table}

\subsubsection{Overall Performance}

\begin{figure*}[!t]
\centering
\includegraphics[width=0.9\textwidth]{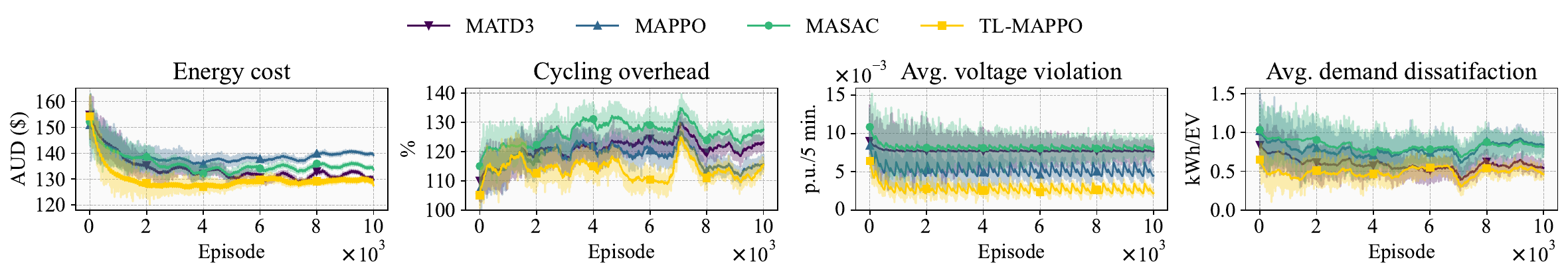}
\vspace{-10pt}
\caption{Learning curves of the developed TL-MAPPO and three baselines.} 
\label{fig:training_curve}
\vspace{-10pt}
\end{figure*}

The training curves in Fig. \ref{fig:training_curve} and the results in Table \ref{Tab:test_results} together show that TL-MAPPO achieves superior learning efficiency, stability, and constraint satisfaction. It yields the lowest daily energy cost (about 125 AUD), whereas baselines remain between 130-150 AUD.
TL-MAPPO also exhibits the smallest voltage violations, lowest unmet EV demand, and the most stable cycling overhead, with consistently narrow confidence intervals. Convergence is faster and smoother across all metrics.
Table \ref{Tab:test_results} further confirms its advantage over 100 episodes: energy cost is reduced by up to 10.7\%, voltage violations by approximately 45\%, and demand dissatisfaction by up to 35\%, with the lowest variance among all methods.

Overall, TL-MAPPO strikes a balanced trade-off between economic operation and grid safety, outperforming MARL baselines under partial visibility. These gains stem from adaptive constraint handling via Lagrangian multipliers that balance reward and safety objectives, and from transformer-assisted observation encoding that captures temporal dependencies for improved long-term control.

\begin{figure*}[!t]
    \centering   
    \renewcommand{\subfigcapskip}{-2pt}
    \begin{minipage}[b]{0.32\textwidth}
        \centering
        \includegraphics[width=\textwidth]{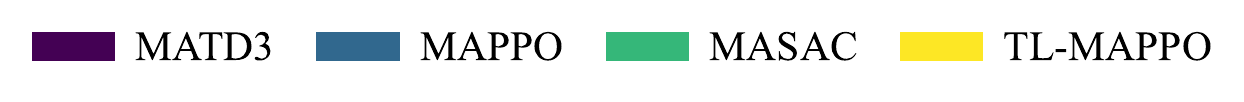}
    \end{minipage}\\
    \vskip -5pt
    \subfigure{
    \includegraphics[width=.21\textwidth]{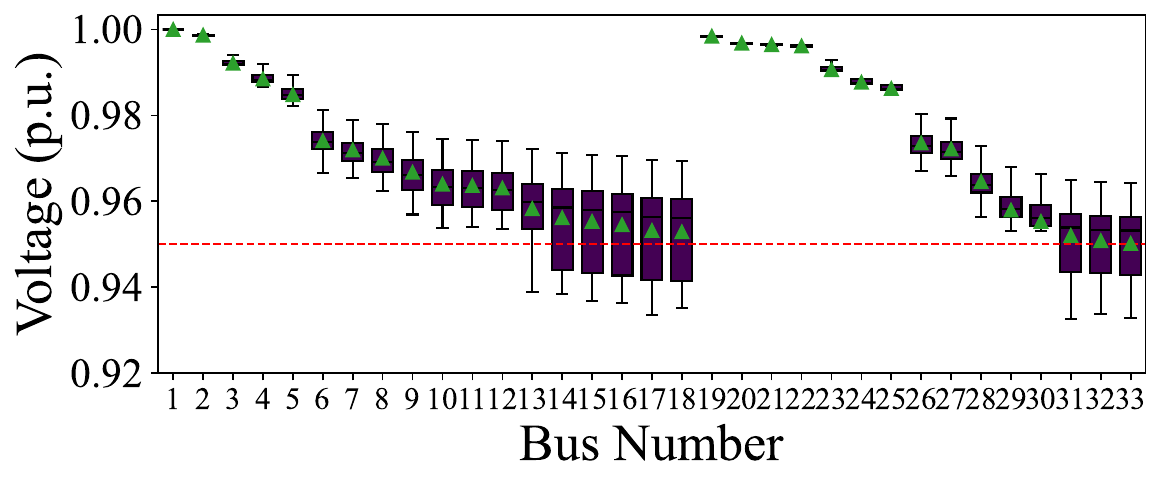}  }
    \hspace{-4mm}
    \subfigure{
    \includegraphics[width=.21\textwidth]{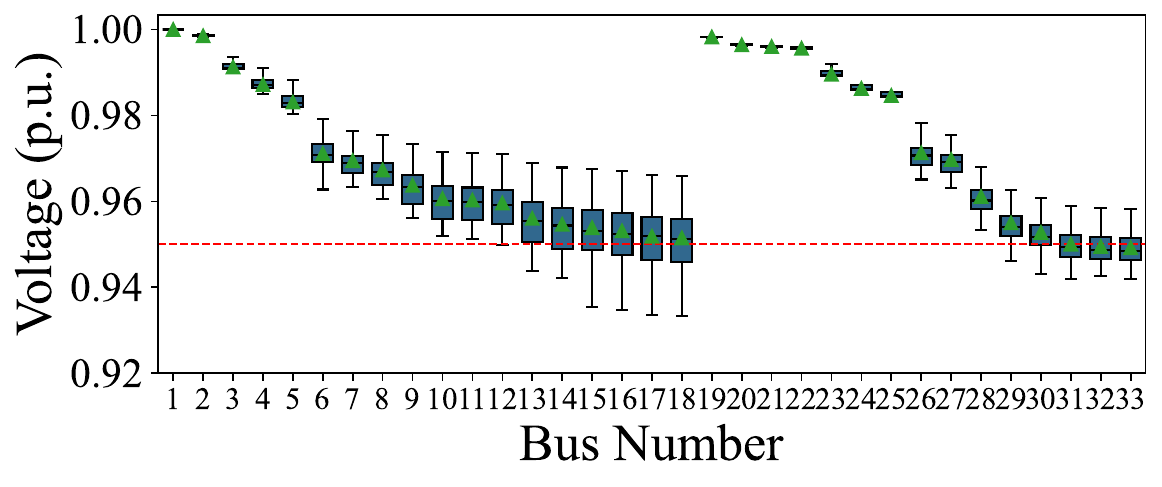}  }
    \hspace{-4mm}
    \subfigure{
    \includegraphics[width=.21\textwidth]{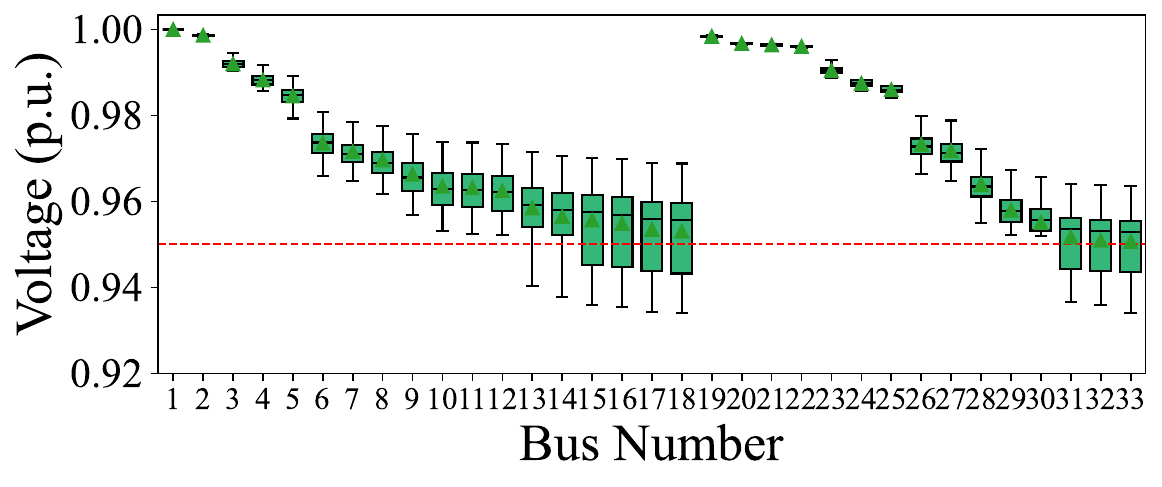}  }
    \hspace{-4mm}
    \subfigure{
    \includegraphics[width=.21\textwidth]{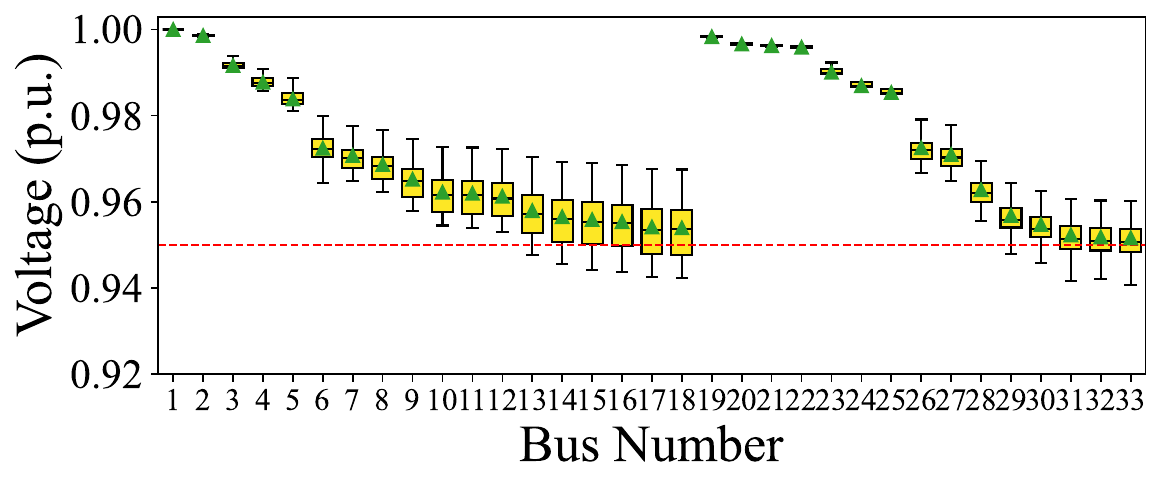}  }
    \vskip -6pt
    \caption{Voltage statistical distributions of buses in one day of the compared methods.}
    \label{Fig: volt_stat}
    \vspace{-10pt}
\end{figure*}

\begin{figure}[!t]
    \centering   
    \renewcommand{\subfigcapskip}{-2pt}
    \subfigure[EV demand, base load, and charging power of EVCS 1]{
    \includegraphics[width=.45\textwidth]{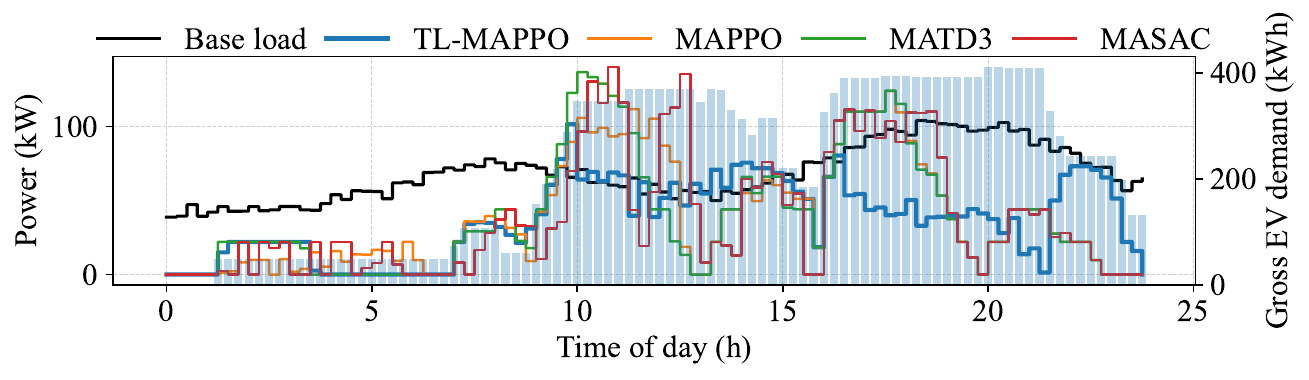}  }
    \vskip -5pt
    \subfigure[Voltage of EVCS 1]{
    \includegraphics[width=.45\textwidth]{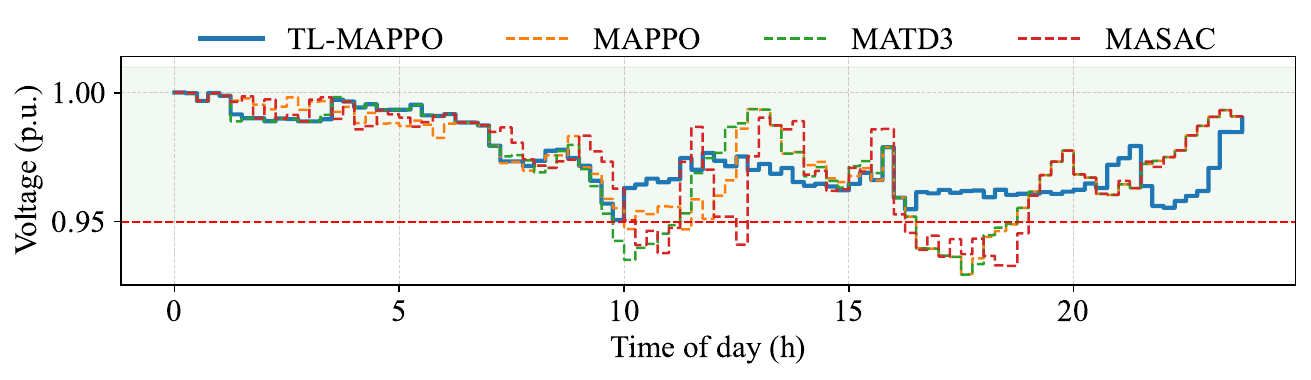}  }
    \vskip -6pt
    \caption{EVCS-level comparison of (a) charging power rate and (b) voltage.}
    \label{Fig: behavior}
    \vspace{-10pt}
\end{figure}

\subsubsection{Voltage Analysis}
Fig.~\ref{Fig: volt_stat} shows the voltage distributions across all buses over a representative day. TL-MAPPO maintains a much tighter and safer voltage range than the baselines, keeping almost all downstream buses above the 0.95 p.u. limit, whereas the other methods exhibit frequent undervoltage, especially at buses 14-33.

Fig.~\ref{Fig: behavior}(a) further compares charging behaviors. TL-MAPPO provides the most stable and grid-aware profile, tracking demand while suppressing charging during peak-load periods. In contrast, MAPPO and MATD3 charge aggressively whenever EVs are present, and MASAC displays strong oscillations.
As shown in Fig.~\ref{Fig: behavior}(b), TL-MAPPO consistently preserves voltages within the safe band, while the other controllers often cause voltages to drop below 0.95 p.u. Overall, TL-MAPPO achieves the most reliable voltage regulation and robust performance across network locations.

\section{Conclusion}\label{Sec:C}
In this paper, we investigated the coordinated EV charging from a VPP perspective under partial distribution network visibility. To address voltage safety, limited network information, and decentralized decision-making, we proposed TL-MAPPO, which integrates Lagrangian-based safety regulation with Transformer-assisted temporal encoding in a multi-agent reinforcement learning framework. Simulation results on a realistic distribution network show that the proposed method reduces voltage violations and operational costs compared with representative MARL baselines while maintaining charging demand satisfaction. Future work will investigate larger-scale VPP deployments and more communication-efficient coordination mechanisms.

\bibliography{reference}
\bibliographystyle{IEEEtran}

\newpage
\appendix

\subsection{Methodological Details}

\subsubsection{Transformer}

To address partial observability in EVCS coordination, a Transformer-based temporal encoder is employed to extract compact representations from historical observations. Specifically, at each decision time step $t$, a temporal observation window is constructed by stacking the local observations defined in Eq. (13) over a fixed horizon, forming a matrix-valued input that captures recent system evolution.

The Transformer encoder consists of multiple stacked self-attention layers operating on this temporal observation window. Multi-head self-attention is used to model heterogeneous temporal dependencies among different input modalities, including electricity prices, PV generation, EV charging states, and local or neighboring voltage measurements. Compared with recurrent architectures, the attention mechanism enables flexible weighting of historical observations across multiple time scales, which is particularly suitable for handling delayed and cumulative effects commonly observed in EV charging and distribution network dynamics. The output of the Transformer is a fixed-dimensional temporal embedding, which is then fed into the decentralized actor networks for action generation.

The main hyperparameters of the Transformer-based temporal encoder are summarized in Table~\ref{tab:transformer_hyper}, including the number of layers, hidden dimension, attention heads, dropout rate, optimizer, and learning rate. These values follow common practice in reinforcement learning and time-series representation learning and are kept fixed across all experiments without extensive hyperparameter tuning, ensuring a computationally lightweight and stable observation-processing module.

\begin{table}[hbpt]
\centering
\caption{Transformer Hyperparameters}
\label{tab:transformer_hyper}
\begin{tabular}{|l|l|}
\hline
\textbf{Hyperparameter} & \textbf{Value} \\
\hline
Number of layers & 6 \\
\hline
Hidden dimension & 256 \\
\hline
Attention heads & 4 \\
\hline
Dropout & 0.1 \\
\hline
Optimizer & Adam \\
\hline
Learning rate & $1\times10^{-4}$ \\
\hline
\end{tabular}
\end{table}

\subsubsection{Overall Algorithm}

As shown in Algorithm \ref{alg:tl_mappo}, the training loop of TL-MAPPO is explicitly outlined, including the Transformer-based observation embedding and Lagrangian update, as provided below to improve clarity and reproducibility.

\begin{algorithm}[hbpt]
\caption{TL-MAPPO: Transformer-assisted Lagrangian MAPPO for Safe EVCS Coordination}
\label{alg:tl_mappo}
\begin{algorithmic}[1]
\REQUIRE Number of agents $K$; rollout length $T$; observation window size $w$; 
policy parameters $\{\theta_k\}_{k=1}^{K}$; centralized critics $\phi_R,\phi_C$; 
Transformer encoder params $\psi$; Lagrangian multiplier $\lambda \ge 0$; 
constraint limit $\bar{C}$; learning rates $\eta_{\theta},\eta_{\phi},\eta_{\psi},\eta_{\lambda}$;
PPO clip $\epsilon$; discount $\gamma$.
\FOR{iteration $=1,2,\dots$}
    \STATE Initialize empty trajectory buffer $\mathcal{D} \leftarrow \emptyset$.
    \STATE \textit{\#Rollout collection}
    \FOR{$t=1$ to $T$}
    \FOR{each agent $k \in \{1,\dots,K\}$} 
            \STATE Observe raw features $\mathbf{o}_{k,t}$ according to Eq.~(13).
            \STATE Construct temporal window $\mathbf{W}_{k,t}=[\mathbf{o}_{k,t-w+1};\dots;\mathbf{o}_{k,t}]$ (pad if $t<w$).
            \STATE Temporal embedding $\hat{\mathbf{o}}_{k,t} \leftarrow g_{\psi}(\mathbf{W}_{k,t})$ \COMMENT{Transformer encoder, Eq.~(16a)}
            \STATE Sample action $\mathbf{a}_{k,t}\sim \pi_{\theta_k}(\cdot \mid \hat{\mathbf{o}}_{k,t})$ \COMMENT{Decentralized actor, Eq.~(17)}
        \ENDFOR
        \STATE Execute joint action $\mathbf{a}_t=(\mathbf{a}_{1,t},\dots,\mathbf{a}_{K,t})$ in environment; receive reward $r_t$ and cost $c_t$.
        \STATE Store $(\{\hat{\mathbf{o}}_{k,t}\}_{k=1}^{K}, \mathbf{a}_t, r_t, c_t)$ into $\mathcal{D}$.
    \ENDFOR

    \STATE Compute returns/advantages $ \hat{A}^R_t, \hat{A}^C_t$ using centralized critics $V_{\phi_R},V_{\phi_C}$ (e.g., GAE), per Eq.~(18).
    \STATE Form Lagrangian advantage $\hat{A}^{\text{Lag}}_t \leftarrow \hat{A}^R_t - \lambda \hat{A}^C_t$ (Eq.~(19a)).

    \STATE \textit{\#PPO updates}
    \FOR{epoch $=1$ to $N_{\text{ppo}}$}
        \FOR{mini-batch $\mathcal{B}\subset \mathcal{D}$}
            \STATE Update each actor $\theta_k$ by maximizing the PPO clipped objective with $\hat{A}^{\text{Lag}}$ (Eq.~(19b)--(19c)).
            \STATE Update critics $\phi_R,\phi_C$ by minimizing squared TD/return losses for reward and cost (Eq.~(18), (20)).
            \STATE Update Transformer encoder $\psi$ jointly via backprop through actor/critic losses (shared encoder).
        \ENDFOR
    \ENDFOR

    \STATE Update Lagrangian multiplier by projected dual ascent:
    \[
    \lambda \leftarrow \Big[\lambda + \eta_{\lambda}\big(\hat{C}-\bar{C}\big)\Big]_+, \quad 
    \hat{C}=\mathbb{E}_{t\in\mathcal{D}}[c_t].
    \]
\ENDFOR
\end{algorithmic}
\end{algorithm}

\subsection{Discussion}

\subsubsection{Communication and Computation}

We consider a high-level coordination architecture between the DSO and the VPP, which is consistent with common abstractions adopted in power system operation studies. In this architecture, the DSO is responsible for monitoring the distribution network and provides the VPP with limited and aggregated network information to support operational decision-making. Such information may include neighborhood-level voltage indicators or aggregated constraint-related signals, rather than full network states or detailed topology.

This abstraction reflects practical considerations regarding data accessibility and privacy in real-world distribution networks. By restricting information exchange to aggregated or local indicators, the framework avoids the need for sharing fine-grained customer-level or network-wide data, thereby preserving privacy-preserving by design. Detailed communication protocols, synchronization mechanisms, and cryptographic privacy-preserving techniques are intentionally abstracted away, as the focus of this work is on control and learning under limited information rather than on communication system design.

From a computational perspective, the proposed framework follows the CTDE paradigm. During training, centralized critics at the VPP aggregate information across EVCSs to evaluate joint system performance and constraint violations. During execution, each EVCS independently runs a lightweight policy using only locally available observations and the learned policy parameters, without requiring real-time communication with other agents or the VPP. This separation enables efficient online operation and aligns with realistic deployment constraints in distribution networks.

\begin{table*}[hbpt]
\centering
\caption{Ablation Study on Temporal Encoding in Lag-MAPPO}
\label{tab:ablation_temporal}
\begin{tabular}{|p{3.5cm}|p{2cm}|p{2cm}|p{3cm}|p{3cm}|}
\hline
Method & 
Energy cost (\$) & 
Cycling overhead (\%) & 
Avg. volt. vio. (p.u./5 min) & 
Avg. demand dissat. (kWh/EV) \\
\hline
Lag-MAPPO (no temporal encoding) &
136.1 $\pm$ 3.6 &
112.8 $\pm$ 2.3 &
4.9 $\pm$ 0.4 $\times 10^{-3}$ &
0.63 $\pm$ 0.05 \\
\hline
Lag-MAPPO + LSTM &
135.2 $\pm$ 3.7 &
111.9 $\pm$ 2.4 &
4.6 $\pm$ 0.4 $\times 10^{-3}$ &
0.61 $\pm$ 0.05 \\
\hline
TL-MAPPO (Transformer) &
\textbf{133.5 $\pm$ 3.4} &
\textbf{110.2 $\pm$ 2.1} &
\textbf{4.2 $\pm$ 0.4 $\times 10^{-3}$} &
\textbf{0.58 $\pm$ 0.05} \\
\hline
\end{tabular}
\end{table*}

\subsubsection{Scalability}

The proposed framework is designed with scalability in mind from an architectural standpoint. As the number of EVCSs increases, communication and computational overhead primarily scale at the VPP side during centralized training, since aggregated information from multiple EVCSs is used to update centralized critics. In contrast, the communication requirement for each individual EVCS remains constant during execution, as agents rely solely on local observations and do not exchange information with one another.

Similarly, computational complexity at the EVCS level is fixed and independent of the total number of stations. Each EVCS executes a decentralized policy with constant-time inference complexity, making the online control cost insensitive to system scale. This property is particularly important for practical deployment, where EVCS controllers may have limited computational resources and strict real-time requirements.

We acknowledge that increasing the number of EVCSs introduces additional challenges beyond computational and communication considerations. In particular, coordination complexity and partial observability may lead to performance degradation as system scale grows, which is a general and well-recognized challenge in large-scale decentralized control and multi-agent reinforcement learning. Addressing these challenges may require more structured coordination mechanisms, such as hierarchical control architectures, clustered agent coordination, or communication-efficient training strategies.

Exploring such extensions, together with systematic evaluation under larger-scale VPP deployments, represents a promising direction for future work. The current study aims to establish a sound and scalable foundation for safe EVCS coordination under realistic information constraints, upon which more advanced large-scale coordination mechanisms can be built.

\subsection{Supplementary Experiments}

We conducted a unified ablation study to isolate the contribution of temporal encoding under the same Lag-MAPPO framework, while keeping the training setup, constraints, and evaluation protocol identical. Specifically, we compared three variants: Lag-MAPPO without temporal encoding, Lag-MAPPO with an LSTM-based encoder, and the proposed TL-MAPPO with a Transformer-based temporal encoder, as summarized in Table~\ref{tab:ablation_temporal}.

The results show a clear performance improvement trend. Introducing temporal encoding already improves performance compared to the non-temporal baseline, indicating that historical context is important under partial observability. Among temporal encoders, the Transformer consistently outperforms the recurrent alternative, achieving lower energy cost, reduced voltage violations, and improved demand satisfaction. These results suggest that the gains of TL-MAPPO do not merely stem from adding temporal memory, but from the Transformer's ability to flexibly capture heterogeneous and long-range temporal dependencies. Overall, the ablation confirms that Lagrangian safety handling enforces constraint compliance, while Transformer-based temporal encoding provides complementary improvements in both safety and economic performance.

\vfill

\end{document}